\documentclass[aps,prb,twocolumn,showpacs,superscriptaddress,floatfix]{revtex4}
\usepackage[utf8]{inputenc}
\usepackage{amsfonts}
\usepackage[intlimits]{amsmath}
\usepackage{bbm,xcolor}
\usepackage{stmaryrd}
\usepackage{graphicx}
\usepackage[bookmarks]{hyperref}

\def\bra#1{\ensuremath{\langle{#1}\vert}}
\def\ket#1{\ensuremath{\vert{#1}\rangle}}

\def\expect#1{\ensuremath{\langle{#1}\rangle}}
\def\abs#1{\ensuremath{\left|{#1}\right|}}
\def\figref#1{\figurename~\ref{#1}}

\newcommand{\Eq}[1]{Eq.~(\ref{#1})}

\renewcommand{\figurename}{Fig.}

\begin{document}

\title{Kondo screening cloud in the single-impurity Anderson model: A DMRG study}

\author{Andreas \surname{Holzner}}
\affiliation{Physics Department, Arnold Sommerfeld Center for Theoretical Physics, and Center for NanoScience,
  Ludwig-Maximilians-Universit\"at M\"unchen, D-80333 M\"unchen, Germany}
\affiliation{Institute for Theoretical Physics C, RWTH Aachen University, D-52056 Aachen, Germany}
\author{Ian P. \surname{McCulloch}}
\affiliation{School of Physical Sciences, University of Queensland, Brisbane, Queensland 4072, Australia}
\author{Ulrich \surname{Schollw\"ock}}
\affiliation{Physics Department, Arnold Sommerfeld Center for Theoretical Physics, and Center for NanoScience,
  Ludwig-Maximilians-Universit\"at M\"unchen, D-80333 M\"unchen, Germany}
\author{Jan \surname{von Delft}}
\affiliation{Physics Department, Arnold Sommerfeld Center for Theoretical Physics, and Center for NanoScience,
  Ludwig-Maximilians-Universit\"at M\"unchen, D-80333 M\"unchen, Germany}
\affiliation{Kavli Institute for Theoretical Physics, Kohn Hall, University of California, Santa Barbara, California
  93106, USA}
\author{Fabian \surname{Heidrich-Meisner}}
\affiliation{Institute for Theoretical Physics C, RWTH Aachen University, D-52056 Aachen, Germany}
\affiliation{Kavli Institute for Theoretical Physics, Kohn Hall, University of California, Santa Barbara, California
  93106, USA}

\date[Date: ]{June 19, 2009}

\begin{abstract}
  A magnetic moment in a metal or in a quantum dot is, at low
  temperatures, screened by the conduction electrons through the
  mechanism of the Kondo effect. This gives rise to spin-spin
  correlations between the magnetic moment and the conduction
  electrons, which can have a substantial spatial extension. We study
  this phenomenon, the so-called Kondo cloud, by means of the density
  matrix renormalization group method for the case of the
  single-impurity Anderson model. We focus on the question whether the
  Kondo screening length, typically assumed to be proportional to the
  inverse Kondo temperature, can be extracted from the spin-spin
  correlations. For several mechanisms -- the gate potential and a
  magnetic field -- which destroy the Kondo effect, we investigate the
  behavior of the screening cloud induced by these perturbations.
\end{abstract}


\maketitle

\section{Introduction}
\label{sec:intro}

The Kondo effect,\cite{Kondo1964} a well-known feature of magnetic
impurity systems, has seen a tremendous renewed interest due to the
realization of quantum dots and nanoscale systems.\cite{Wiel2002} The
existence of Kondo correlations at low temperatures $T$ has been firmly
established in numerous experiments on quantum
dots,\cite{Goldhaber-GordonShtrikmanMahaluAbusch-MagderMeiravKastner1998}
molecules,\cite{LiangShoresBockrathLongPark2002} and carbon
nanotubes.\cite{ZhengLuGuMakarovskiFinkelsteinLiu2002} The
interaction of an impurity spin with itinerant electrons, causing the
Kondo effect, manifests itself in spatially extended spin-spin correlations
-- the Kondo screening cloud. These correlations have been extensively
studied in theory\cite{GubernatisHirschScalapino1987,BarzykinAffleck1996,SorensenAffleck1996,AffleckSimon2001,SorensenAffleck2005,HandKrohaMonien2006,CostamagnaGazzaTorioRiera2006,Borda2007a,AffleckBordaSaleur2008,PereiraLaflorencieAffleckHalperin2008}
and many proposals for experimentally measuring the Kondo screening
cloud have been put
forward.\cite{AffleckSimon2001,HandKrohaMonien2006,AffleckBordaSaleur2008,PereiraLaflorencieAffleckHalperin2008}
Also, several studies have emphasized the emergence of mesoscopic
fluctuations on finite systems, and the existence of even-odd effects
in the Kondo cloud when computed from a lattice
model.\cite{SorensenAffleck1996,AffleckSimon2001,SimonAffleck2003,ThimmKrohaDelft1999,HandKrohaMonien2006}
While there has been experimental progress toward the measurement of the
Kondo
cloud,\cite{MadhavanChenJamnealaCrommieWingreen1998,ManoharanLutzEigler2000}
the detection of the spin-spin correlations has proven to be highly
challenging and has not been accomplished so far.  Depending on the
Kondo temperature $T_K$, the Kondo cloud can have a significant
extension of $\sim 1 \mu m$.\cite{Borda2007a}

In our work, we examine the spin-spin correlations in a real-space model, the single-impurity Anderson model (SIAM) that
includes charge fluctuations, using the density matrix renormalization group method
(DMRG).\cite{white:dmrg1,white:dmrg2,Schollwock2005} We address two main questions: first,  we compute the spin-spin
correlations between the impurity spin and the conduction electrons at particle-hole symmetry and discuss how the Kondo
screening length $\xi_{K}$ can be directly extracted from such data. 
To that end, we discuss several ways of collapsing
spin-spin correlations calculated for different Kondo temperatures onto a  universal curve.  
In this analysis, we employ ideas suggested by Gubernatis {\it et al.}\cite{GubernatisHirschScalapino1987} that
have also been used in previous DMRG studies of the Kondo cloud
problem.\cite{HandKrohaMonien2006,CostamagnaGazzaTorioRiera2006} We find that from chains of
about $L=500$ sites, suitable measures for  the $L=\infty$ screening length can be extracted for Kondo temperatures of
$k_{B}T_{K}/\Gamma\sim 1 \cdot 10^{-3}$ ($\Gamma$ is the tunneling rate). Knowledge of the universal curve further
allows us to estimate $\xi_K$ even for Kondo temperatures for which the accessible system sizes are too small to host
the full Kondo cloud. As a main result of our analysis, we find that our measures of $\xi_K$ extracted from the
spin-spin correlations have the same functional dependence on model parameters as $\xi_K^0$,
\begin{equation}
  \xi_K^0=\hbar v_F/T_K\,,\label{eq:tk_xi}
\end{equation}
at particle-hole symmetry  ($v_F$ is the Fermi velocity in the leads, we adopt $k_{B} = 1$ throughout the rest of this
work). The screening length $\xi_{K}^{0}$ governs the finite-size scaling of local quantities such as the polarization
or the magnetic moment.\cite{SorensenAffleck1996}

Second, we consider several mechanisms that destroy Kondo correlations, namely a gate voltage and a magnetic field
applied to the quantum dot. We study the changes  in the screening length induced by a variation in these parameters. We
argue that computing the magnetic field dependence of the screening length provides a means of extracting the Kondo
temperature.

The emergence of an exponentially small energy scale in the Kondo problem, namely $T_{K}$, restricts any real-space
approach with respect to the Kondo temperatures that can be accessed. A powerful framework was introduced by
Wilson\cite{wilson:nrg} in the form of the numerical renormalization group (NRG)
method,\cite{wilson:nrg,BullaCostiPruschke2008} which is explicitly tailored toward the Kondo problem. This is achieved
through the introduction of a logarithmic energy discretization that allows the Kondo scale to be resolved but loses
real space information. Recently, an NRG method has been developed to access spatially resolved
quantities,\cite{Borda2007a,BordaGarstKroha2008,AffleckBordaSaleur2008} extending some older NRG calculations for
spatially dependent correlation functions.\cite{ChenJayaprakashKrishnamurthy1992} Using the more recent NRG
approach,\cite{Borda2007a} the spin correlations between the impurity and the sites in the leads have been computed for
the Kondo model, and it has been shown that at the Kondo screening length $\xi_K^0$, the envelope of the correlations
crosses over from a $1/x^2$ decay at distances $x< \xi_K^0$ to a $1/x$ decay at distances $x>\xi_K^0$, where $x$ denotes
the distance between the impurity and a site in the leads.

Comparing NRG and DMRG, first, there are technical differences between DMRG and NRG with respect to how the spin-spin correlations
$\expect{\vec{S}_{i}\cdot\vec{S}_{j}}$ ($\vec{S}_i$ denotes a spin-1/2 operator at site $i$) are  obtained. NRG requires a
separate run for each pair of indices $(i,j)$, whereas DMRG  operates directly on real-space leads. That way, after
calculating the ground state for a system of a given length, all correlations can be evaluated in a single run. While
the use of real-space chains is restricted to one dimension, which is the case of interest in our work, NRG in principle
works for higher dimensions too. Second, using DMRG, we can gain direct and easy information on the finite-size scaling
of spin-spin correlations, which we heavily exploit in our analysis. Most importantly, DMRG can also be applied to
quantum-impurity problems with interacting leads\cite{CostamagnaGazzaTorioRiera2006} that NRG is not designed for.

DMRG has previously been used to study the Kondo cloud in several papers, for both the single-impurity Anderson model
\cite{HandKrohaMonien2006} and the Kondo model.\cite{SorensenAffleck2005,SorensenAffleck1996} In
Ref.~\onlinecite{HandKrohaMonien2006} by Hand {\it et al.}, in particular, an interesting relation between the screening length as extracted
from the spin correlations and the weight of the Kondo resonance has been discussed. Our study extends the DMRG
literature as we consider the mixed-valence regime, the effect of a magnetic field, and we discuss and demonstrate the
universal scaling of spin-spin correlations for a wide range of parameters.  Moreover, in the absence of a magnetic
field, we exploit the SU(2) symmetry of the model in the spin  sector in the DMRG simulations, which we find is crucial
for efficiently obtaining reliable numerical results.

Besides the conceptual interest in understanding the scaling properties of the Kondo screening length with both system
size and Kondo temperature, our results are relevant to gauge the range of validity of numerical approaches for
calculating the conductance of nanostructures that employ a real-space representation of the leads such as
time-dependent DMRG simulations of transport in the single-impurity Anderson
model.\cite{Al-HassaniehFeiguinRieraBuesserDagotto2006,KirinoFujiiZhaoUeda2008,Heidrich-MeisnerFeiguinDagotto2009}
Moreover, the approaches discussed here to extract the screening length could be applicable to more complex geometries
in a straightforward way, for instance, to multichannel and/ or multidot problems.

Our work is organized as follows. In Sec.~\ref{sec:model}, we introduce  our model and define the quantities of
interest. In Sec.~\ref{sec:spin-correlations}, the spin-spin correlations constituting the Kondo cloud are investigated
and we demonstrate how to extract the value of the Kondo screening length $\xi_{K}$ from the spin correlation data,
making use of the universal finite-size scaling behavior of $\xi_{K}$. We proceed with a discussion of the behavior of
the screening length upon driving the system away from the Kondo point via a gate potential, presented in
Sec.~\ref{sec:gate-potential}, and then turn to the case of a magnetic field in Sec.~\ref{sec:field}. We conclude with a
summary, Sec.~\ref{sec:summary}, while technical detail on the method and computations are given the Appendix.

\section{Model}
\label{sec:model}
We model a quantum dot coupled to a lead by the single-impurity Anderson model, describing the lead by a tight-binding
noninteracting chain. This constitutes a one-channel problem:
\begin{equation}
  \begin{split}
    H &= \sum_{\sigma = \shortuparrow,\shortdownarrow} \epsilon_{d} n_{d\sigma} + B S^{z}_{d} + U n_{d\shortuparrow}
    n_{d\shortdownarrow}\\
    &- t\sum_{\sigma}\sum_{i=1}^{L-1} \left( c_{i\sigma}^{\dagger} c_{(i+1)\sigma}^{\phantom{*}} + {\rm h.c.} \right)\\
    &- \sum_{\sigma} \sqrt{2} t' \left( c_{1\sigma}^{\dagger} d_{\sigma}^{\phantom{*}} + {\rm h.c.} \right).
  \end{split}
\end{equation}
$c_{i \sigma}$ annihilates an electron with spin $\sigma = \shortuparrow,\shortdownarrow$ on site $i$, $d_{\sigma}$
annihilates an electron with spin $\sigma$ on the dot, and $n_{d\sigma} =
d^{\dagger}_{\sigma}d^{\phantom{*}}_{\sigma}$. The spin operators at any site are given by $S^{a}_{i} =
c^{\dagger}_{is}{\sigma^{a}_{ss'}}c^{\phantom{*}}_{is'}/2$, where $\sigma^{a}$ are the Pauli matrices
($a=x,y,z$). $\epsilon_{d}$ denotes the gate potential and $B$ denotes the magnetic field applied to the dot, $U$
denotes the strength of the Coulomb interaction on the quantum dot, $t'$ denotes the hopping of the dot levels to the
first site in the lead, $t$ denotes the hopping within the lead. The width of the dot level due to the hybridization
with the lead is given by $\Gamma = 2{t'}^{2}/t$.

In the absence of a magnetic field, this model has a spin SU(2) symmetry. In our analysis, we calculate the ground state
of this system via DMRG using an implementation\footnote{While we use a matrix-product-states-based implementation of
  DMRG,\cite{McCulloch2007} for the problem studied here, equivalently good results can be obtained with standard DMRG
  codes that exploit sufficiently many good quantum numbers.} exploiting the SU(2) symmetry which greatly improves the
efficiency \cite{McCullochGulacsi2002,McCulloch2007} (see the Appendix for more detail). A typical run for $L=500$
sites with $m=1500$ states took about 60 h on a 2.6~GHz Opteron CPU.

All simulations, irrespective of $\epsilon_d$, are performed at half-filling of the full system. As the Kondo scale depends
exponentially on $U/\Gamma$, while in a real-space representation of the leads,  the energy resolution is proportional
to ${1}/{L}$, we restrict our analysis to the intermediate values of ${U}/{\Gamma}$. The tradeoff for these limitations is
that it is straightforward to calculate spin correlators, as outlined below [see Eq.~\eqref{eq:spin-int-corr}].

Throughout this work, we use chains with an overall {\it even} number of sites. It is well-known that there are
significant even-odd effects in impurity problem of this
kind.\cite{SorensenAffleck1996,AffleckSimon2001,SimonAffleck2003,ThimmKrohaDelft1999,HandKrohaMonien2006} Earlier work
(see, {\it e.g.}, Ref.~\onlinecite{heidrichmeisner09a}), suggests that the convergence with system size toward a Kondo
state is much faster on chains with an even number of sites. We thus work in singlet subspaces.

\section{Spin-spin correlations and Kondo screening length at $\epsilon_{d} = -U/2$}
\label{sec:spin-correlations}

In this section, we present our results for the spin-spin-correlation function at particle-hole symmetry and
we discuss two ways of collapsing the data, allowing for a determination of the Kondo screening length.
\begin{figure}[t]
  \includegraphics[width=\linewidth]{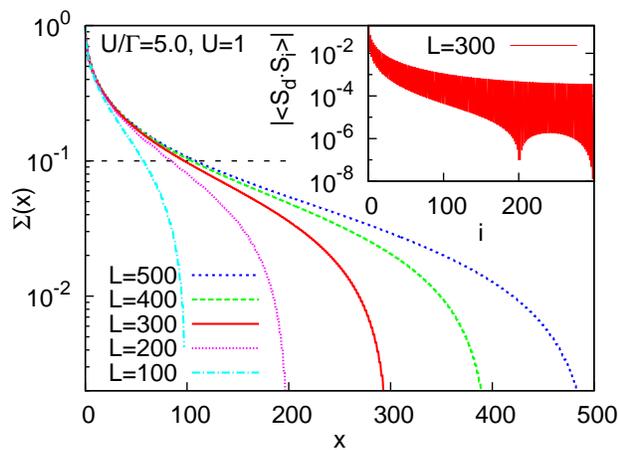}
  \caption{(Color online) Integrated spin-spin correlations $\Sigma(x)$ [from \Eq{eq:spin-int-corr}] for systems of
    different sizes at $U=1, \Gamma=0.20$ and $\epsilon_d=-U/2$. As an example, the threshold of $0.1$ that we use in
    Eq.~\eqref{eq:xi_09} to extract $\xi_{0.9}$ is indicated by the dashed horizontal line. As an illustration of the
    typical raw data, we show the absolute value of the spin-spin correlations $|\expect{\vec{S}_{d}\cdot\vec{S}_{i}}|$
    for $L=300$ in the inset.}
  \label{fig:spin-correlations}
\end{figure}
In order to investigate the behavior of the Kondo screening length, we shall study the following integrated
spin-correlation function,
\begin{equation}
  \label{eq:spin-int-corr}
  \Sigma(x) = 1 + \sum_{i=1}^{x}\frac{
    \expect{\vec{S}_{d}\cdot\vec{S}_{i}}}{\expect{\vec{S}_{d}\cdot\vec{S}_{d}}} , 
\end{equation}
to be evaluated in the singlet subspace of the total spin
$\vec{S}_{\mathrm{tot}}=\vec{S}_d+\sum_{i=1}^{L-1}\vec{S}_{i}$, and under the assumption that $\langle
\vec{S}_d^2\rangle\not= 0$ ($x$ is given in units of the lattice constant). This definition is motivated by the
following convenient properties: (i) the decay of $\Sigma (x)$ with $x$ characterizes the extent to which the total spin
of chain sites one to $x$ is able to screen the spin on the impurity level, {\it i.e.}, the extent to which $\sum_{i =
  1}^x \vec S_i$ has, crudely speaking, ``become equal and opposite'' to $\vec S_d$. (ii) When the sum includes the
entire chain, we always have $\Sigma(L-1) = 0$. This follows by noting that in  the subspace with zero total spin, where
$\langle \vec{S}^2_{\mathrm{tot}}\rangle = 0$, we have $\langle \vec{S}_d^2\rangle=
\langle(\sum_{i=1}^{L-1}\vec{S}_i)^2\rangle$,  and hence also $\langle \vec{S}^2_{\mathrm{tot}}\rangle = 2
\expect{\vec{S}_{d}\cdot\vec{S}_{d}} \Sigma(L-1)$. (iii) The correlator is normalized to $\Sigma (0) = 1$. (iv) In the
absence of a magnetic field, $\Sigma(x)$ is SU(2) invariant, such that this symmetry can be exploited in our
numerics. In the presence of a magnetic field, we shall use a symmetry-broken version, replacing
$\expect{\vec{S}_{d}\cdot\vec{S}_{i}}/\expect{\vec{S}_{d}\cdot\vec{S}_{d}}$ by $(\expect{{S}^z_{d} {S}^z_{i}} -
\expect{S^{z}_{d}}\expect{S^{z}_{i}})/(\expect{S^z_{d} S^z_{d}} - \expect{S^{z}_{d}}^{2})$.

As an example, the inset of \figref{fig:spin-correlations} shows a DMRG result for the absolute value of the bare
spin-spin correlator $\expect{\vec{S}_d \cdot \vec{S}_i}$. The feature at $i\sim 200$ is a simple effect of the open
boundary conditions. The spin correlations for $i$ smaller than a certain value (here roughly $i\sim 200$) oscillate
between negative and positive, while beyond a certain point, all $\expect{\vec{S}_d \cdot \vec{S}_i}$ become
positive. This feature at $i\sim 200$ precisely appears at the site where this happens, {\it i.e.}, where
$\expect{\vec{S}_d \cdot \vec{S}_i}$ with $i$ even changes its sign and, as a consequence, the correlator passes
arbitrarily close through zero. Summing up the correlator according to Eq.~\eqref{eq:spin-int-corr} yields $\Sigma(x)$,
plotted in the main panel.

The notion of a screening length is based on the premise that the decay of $\Sigma(x)$ follows a universal form
characterized by a single length scale, $\xi_K$, as long as this scale is significantly shorter than the system size,
$\xi_K \ll L$.  (According to the expectation that $\xi_K \propto \hbar v_F / T_K$, this condition is equivalent to the
following statement: perfect spin screening in a system of finite size $L$ can only be achieved if the level spacing,
which scales like $\hbar v_F / L$, is smaller than $T_K$.)  Whenever this condition is not met, the shape of the decay
of $\Sigma(x)$ with $x$ deviates from its universal form once $x$ becomes large enough such that the finite system size
makes itself felt [via the boundary condition $\Sigma(L-1) = 0$].  To extract $\xi_K$ from DMRG data obtained for
finite-sized systems, we thus need a strategy for dealing with this complication.  Below, we shall describe  two
different approaches that accomplish this, both involving a scaling analysis.

To check whether the screening length  obtained using either of the two scaling strategies conforms to the theoretical
expectations, we shall check whether its dependence on the parameters $U$, $\Gamma$, and $\epsilon_d$ agrees with that of
the length scale $\xi_{K}^{0} = \frac{\hbar v_{F}}{T_{K}}$ [Eq.~\eqref{eq:tk_xi}]. Using the known form of the Kondo
temperature $T_K$ for the Anderson model,\cite{Haldane1978a,ZitkoBoncaRamsakRejec2006} this dependence is given by:
\begin{equation}
  \label{eq:kondo-scale}
  \xi_K^0 \equiv   \frac{ \hbar v_F}{
    \sqrt{U\Gamma}} \exp\left[\frac{\pi\abs{\epsilon_{d}}
      \abs{\epsilon_{d}+U}}{2U\Gamma}\right]\,.
\end{equation}      
We shall indeed find a proportionality of the form $\xi_K = p\,\xi_K^0$, where the numerical prefactor $p$ reflects the
fact that the definition of $T_K$ involves an arbitrary choice of a prefactor on the order of one. We emphasize,
however, that our determination of $\xi_K$ will be carried out without invoking \Eq{eq:kondo-scale}; rather, our results
for $\xi_K$ will turn out to confirm Eq.~\eqref{eq:kondo-scale} \emph{a posteriori}. In the present section we shall
focus on the symmetric Anderson model ($\epsilon_d = -U/2$) at zero magnetic field, considering more general cases in
\ref{sec:gate-potential}.

\subsection{Scaling collapse of $\Sigma (x)$}
\label{sec:scaling-sigma}

\begin{figure}[t]
  \includegraphics[width=\linewidth]{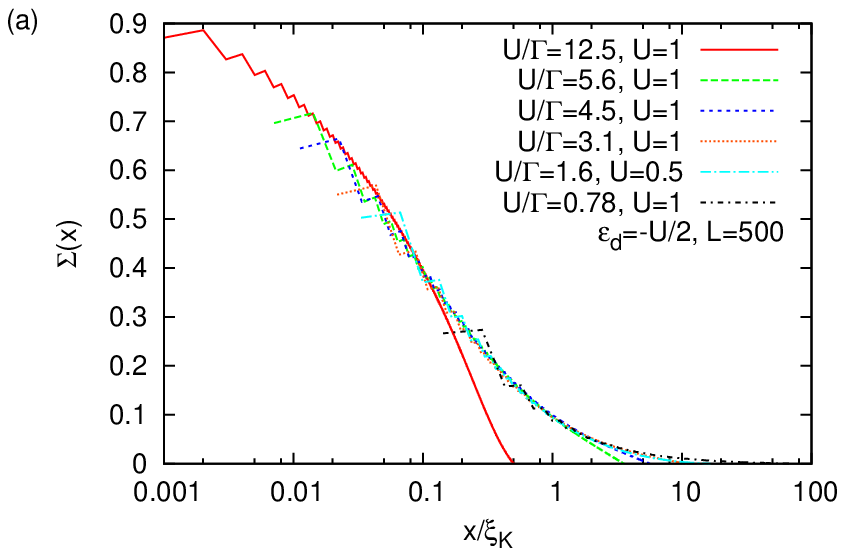}\\
  \includegraphics[width=\linewidth]{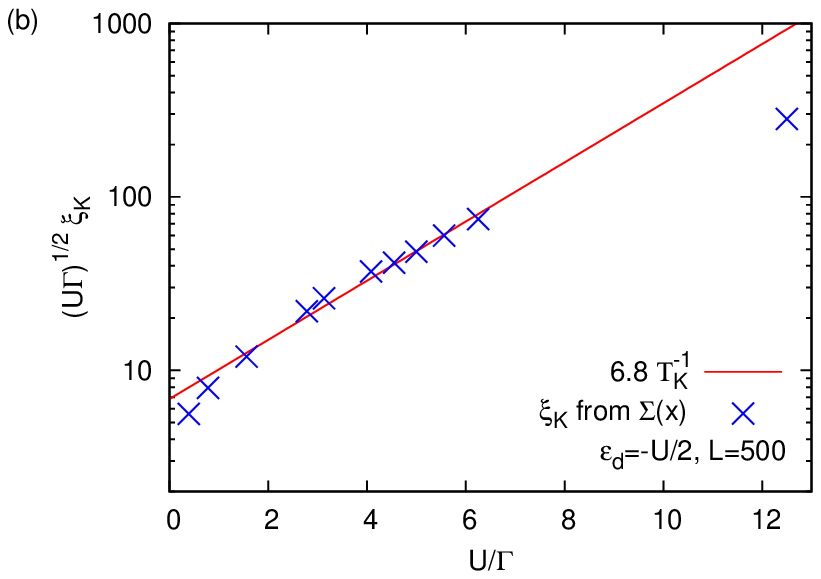}%
  \caption{(Color online) (a) Rescaled integrated spin-spin correlations $\Sigma(x)$, collapsed onto a universal curve
    via suitable choices of $\xi_{K} $. (b) Comparison of the $U$ and $\Gamma$ dependence of $\xi_{K}$ and $\xi_K^0$
    [from \Eq{eq:kondo-scale}]: $\sqrt{U\Gamma} \xi_{K} $ (symbols) and $\sqrt{U \Gamma} p\,T_{K}^{-1}$ (lines) plotted
    vs $U/\Gamma$, using $p$ as the fitting parameter (resulting in $p = 6.8$).}
  \label{fig:sigma-fit}
\end{figure}

The first way of extracting the screening length is to plot $\Sigma(x)$ versus $x/\xi_{K}$, where $\xi_{K}$ is treated
as a fitting parameter, to be chosen such that all the curves collapse onto the same scaling curve [see
\figref{fig:sigma-fit}].
When attempting to collapse the $\Sigma(x)$ data, one faces two issues. First, the $\Sigma(x)$ data are nonmonotonic in
$L$, due to the fact that the sign of $\expect{\vec S_d \cdot \vec S_i}$ oscillates, and for curves scaled by different
values of $\xi_{K}$, the oscillations are stretched by different amounts on a semilog plot. This introduces some
``noise'' to the $\Sigma(x)$ curves, making it somewhat difficult to decide when the scaling collapse is
optimal. Second, for some parameter combinations, the condition $\xi_K \ll L$ is not met, and therefore, perfect scaling
cannot be expected for all the curves.

These issues can be dealt with by a two-step strategy: (i) we start with the curves, which collapse the best, namely,
those with the smallest $U/\Gamma$ ratios. These yield the smallest $\xi_{K}$ values and hence satisfy the condition
$\xi_{K} < L$ required for good scaling well enough such that the shape of the universal scaling curve can be
established unambiguously (to the extent allowed by the aforementioned noise). (ii) We then proceed to larger ratios of
$U/\Gamma$, which yield larger $\xi_{K}$, and adjust $\xi_{K}$ such that a good collapse of $\Sigma(x)$ vs.\ $x/\xi_{K}$
onto the universal curve is achieved in the regime of small $x/\xi_{K}$, where finite-size effects are not yet
felt. Thus, knowledge of the universal scaling curve allows $\xi_{K}$ to be extracted even when the condition $\xi_{K}
\ll L$ is not fully met.

The result of such a scaling analysis is shown in \figref{fig:sigma-fit}(a). A universal scaling curve can clearly be
discerned, with deviations from scaling evident in the curves with large $U/\Gamma$, as expected. Moreover,
\figref{fig:sigma-fit}(b) shows that the results for $\xi_K$ extracted from $\Sigma(x)$ scaling agree rather well with
the parameter dependence expected from \Eq{eq:kondo-scale} for $p/(\hbar v_{F})\cdot\xi_{K}^{0}$ (with a prefactor of
$p=6.8$), provided that $U/\Gamma \gtrsim 2$. For smaller $U/\Gamma$, no well-defined local moment will form and the
premise for \Eq{eq:kondo-scale} no longer holds.

\subsection{Scaling collapse of $\xi_a (L)$} 
\label{sec:xia}
A second strategy for extracting the screening length, following
Refs.~\onlinecite{GubernatisHirschScalapino1987,HandKrohaMonien2006}, and~\onlinecite{CostamagnaGazzaTorioRiera2006}, is
to determine the length, say $\xi_a$, on which the integrated spin-correlation function $\Sigma(x)$ has dropped by a
factor of $a$ of its $x=0$ value (for instance, \ $a = 0.9$ would signify a 90\% screening of the local spin). Thus, we
define
\begin{equation}
  \label{eq:xi_09}
  \xi_{a}(L) = \min \left\{ x; \Sigma(x) \leq 1-a \right\} \,.
\end{equation}
The argument of $\xi_a(L)$ serves as a reminder that this length depends on $L$, since the boundary condition
$\Sigma(L-1) = 0$ always enforces perfect screening for $x=L$. However, once the system size becomes sufficiently large
$(L > \xi_K)$ to accommodate the full screening cloud, $\xi_a(L)$ approaches a limiting value, to be denoted by $\xi_a$
[shorthand for $\xi_a(\infty)$], which may be taken as a measure of the true screening length $\xi_K$. This is
illustrated in the main panel of \figref{fig:spin-correlations} for $a = 0.9$: as $L$ increases, the $x$-values, where
the $\Sigma(x)$ curves cross the threshold $1-a = 0.1$ (horizontal dashed line), tend to a limiting value. This limiting
value, reached in \figref{fig:spin-correlations} for $L>300$, defines $\xi_{0.9}$.

\begin{figure}[t]
  \includegraphics[width=\linewidth]{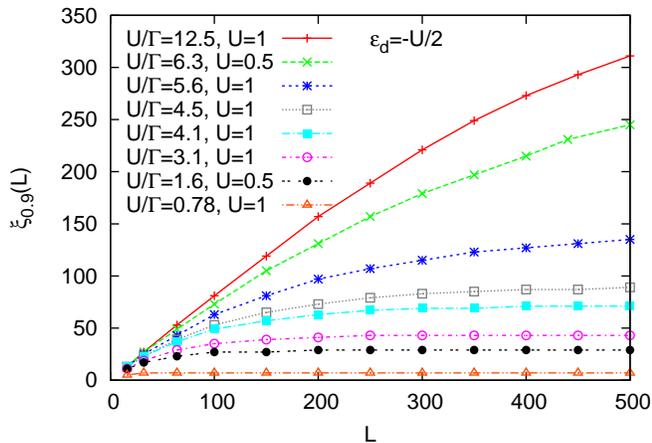}
  \caption{(Color online) System size dependence of $\xi_{0.9}(L)$ for $\epsilon_d=-U/2$. Points represent numerical
    data; lines serve as guides to the eyes.}
  \label{fig:xi-systemsize}
\end{figure}
Figure~\ref{fig:xi-systemsize} shows the $L$ dependence of $\xi_{0.9} (L)$ for several values of ${U}/{\Gamma}$ ranging
from 0.4 to 12.5, and system sizes up to $L=500$.  We observe that $\xi_{0.9}(L)$ reaches its limiting value for small
ratios of $U/\Gamma$, which produce $\xi_{0.9}$ values smaller than $L=500$.  For larger values of $U/\Gamma$, however,
$\xi_{0.9} (L)$ does not saturate, implying that for these parameters, the true screening length is too large to fit
into the finite system size.\footnote{We note that by definition $\xi_{a}(L)$ is only accurate up to one lattice
  constant. As a consequence, very small changes in $\expect{\vec{S}_{d}\cdot\vec{S}_{i}}$ may cause a change in
  $\xi_{a}(L)$ by one. This can be seen in, {\it  e.g.}, the data for $U=1$, $\Gamma=0.22$  from
  \figref{fig:xi-systemsize} (open squares), where $\xi_{0.9}(L)$ is very close to convergence in $L$ but still increases
  between $L=450$ and $L=500$ by one.}

\begin{figure}[t]
  \centerline{\includegraphics[width=\linewidth]{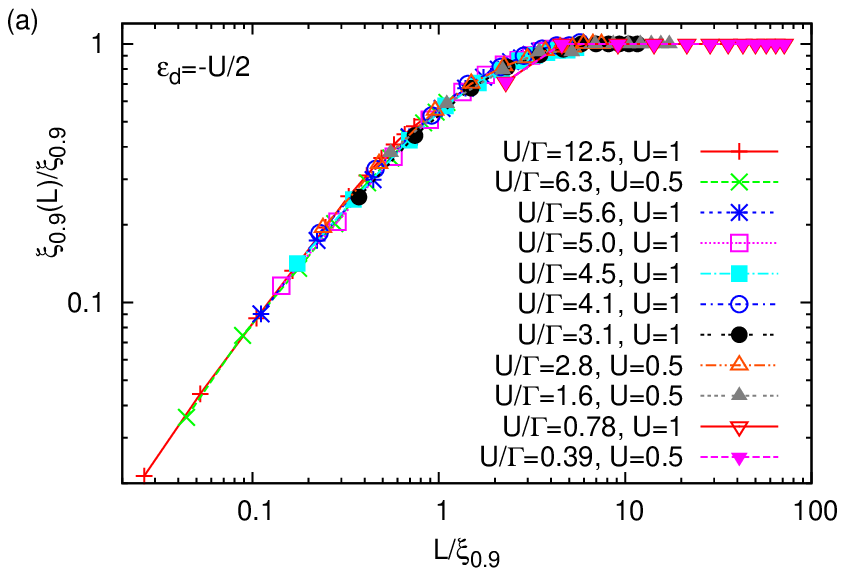}}
  \centerline{\includegraphics[width=\linewidth]{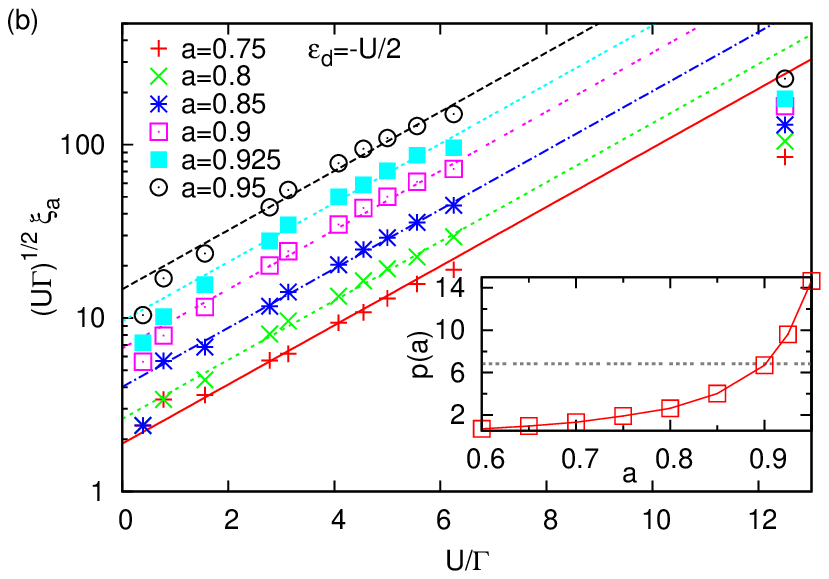}}
  \caption{(Color online) Results of a $\xi_a(L)$ scaling analysis for $\epsilon_d=-U/2$. (a) Scaling collapse of
    $\xi_{0.9}(L)/\xi_{0.9}$ vs $L/\xi_{0.9} $, obtained by the two-step scaling strategy described in the text in
    Sec.~\ref{sec:xia}. (b) Comparison of the $U$ and $\Gamma$ dependence of $\xi_a$ and $\xi_K^0$ [from
    \Eq{eq:kondo-scale}] for several values of $a$: $\sqrt{U \Gamma} \xi_a$ (symbols) and $\sqrt{U \Gamma} p(a)\,
    T_{K}^{-1}$ (lines) plotted vs $U/\Gamma$, using the fit parameters $p(a)$ shown in the inset (squares). The dotted
    line in the inset indicates the prefactor $p=6.8$ obtained from the $\Sigma (x)$ scaling analysis of
    \figref{fig:sigma-fit}(b).}
  \label{fig:xi-fit}
\end{figure}

Nevertheless, it is possible to extract the true screening length in the latter cases as well, by performing a two-step
finite-size scaling analysis: (i) for those parameters $U/\Gamma$ for which $\xi_{a}(L)$ has already saturated on a
finite system, we set $\xi_{a} = \xi_a (L=500)$, and plot $\xi_a (L)/\xi_a$ vs $L/\xi_a$. This collapses all such curves
onto a universal scaling curve. For larger $U/\Gamma$, we rescale the $\xi_a (L)$ curves in a similar fashion, but now
using $\xi_a$ as a fit parameter, chosen such that the rescaled curves collapse onto the universal curve determined in
step (i). As shown in \figref{fig:xi-fit}(a) for $a = 0.9$, this strategy produces an excellent scaling collapse for all
combinations of $U$ and $\Gamma$ studied here.

The above procedure requires the threshold parameter $a$ to be fixed arbitrarily. Qualitatively, one needs a large $a$
to capture most of the correlations {\it i.e.}, $\xi_a(L\to \infty) \sim \xi_K$, yet $a$ ought not to be too close to
one to avoid boundary effects in the results. Technically, the calculation of $\xi_a$ is much easier the smaller $a$ is,
as less correlators $\expect{\vec{S}_i\cdot\vec{S}_{\mathrm{dot}}}$ that are of a small numerical value  need to be
computed to high accuracy (see also the discussion in the Appendix). For instance, at $U/\Gamma=5$ and $L=500$,
$\xi_{0.9} \approx 112$ sites, while $\xi_{0.75} \approx 29$ sites.
  
We have carefully analyzed the qualitative dependence of our analysis on the threshold $a$. First, the universal
scaling behavior in $\xi_{a}(L)/\xi_{a}$ is seen for $a > 0.6$. Using too small a value for $a$ ignores the long-range
behavior of $\Sigma(x)$. Qualitatively, $\xi_{a}$  needs to be close to the point, where the decay of the envelope of
spin-spin correlations changes from a power law with $1/x$ to $1/x^2$ (see Fig.~2 in
Ref.~\onlinecite{Borda2007a}). Second, it turns out that different choices of $a$ produce values of $\xi_a $ that
differ only by a ($U$-independent and $\Gamma$-independent) prefactor $p(a)$, as illustrated in \figref{fig:xi-fit}(b)
(symbols). In particular, for $U/\Gamma \gtrsim 2$, all $\xi_{a}$ follow the same functional dependence on the
parameters $U$ and $\Gamma$, satisfying the relation
\begin{equation}
  \label{eq:xi-a}
  \xi_a  = \frac{p(a)}{\hbar v_{F}} \, \xi_K^0
\end{equation}
expected from Eq.~(\ref{eq:kondo-scale}) (lines in \figref{fig:xi-fit}). It is obvious that $\xi_a$ yields an {\it
  upper} bound to $\xi_K^0$ since $p(a) > 1$ for all choices of $a$.

The only exceptions are the data points at $U/\Gamma = 12.5$, for which $\xi_a $ is too large in comparison to $L=500$
to yield reliable results. The latter are thus excluded when fitting the $\xi_a $ data to determine the best values for
$p(a)$, shown in the inset of \figref{fig:xi-fit}.
 
The inset includes the prefactor $p = 6.8$ (horizontal dotted line) obtained in Sec.~\ref{sec:scaling-sigma}, from
\figref{fig:sigma-fit}, via a scaling analysis of $\Sigma(x)$ (which has the advantage of not involving any arbitrarily
chosen threshold). Evidently, $p=6.8$ is rather well matched  by $p(0.9) \simeq 6.7$, implying that the two alternative
scaling strategies explored above, based on $\Sigma(x)$ and $\xi_a(L)$, yield essentially identical screening lengths
for $a=0.9$. For the remainder of this paper, where we consider $\epsilon_d \neq -U/2$ or $B \neq 0$, we shall thus
determine the screening length by employing $\xi_{0.9} (L)$ scaling, which is somewhat more straightforward to implement
than $\Sigma(x)$ scaling.

\section{Gate potential}
\label{sec:gate-potential}

\begin{figure}[ht]
  \includegraphics[width=\linewidth]{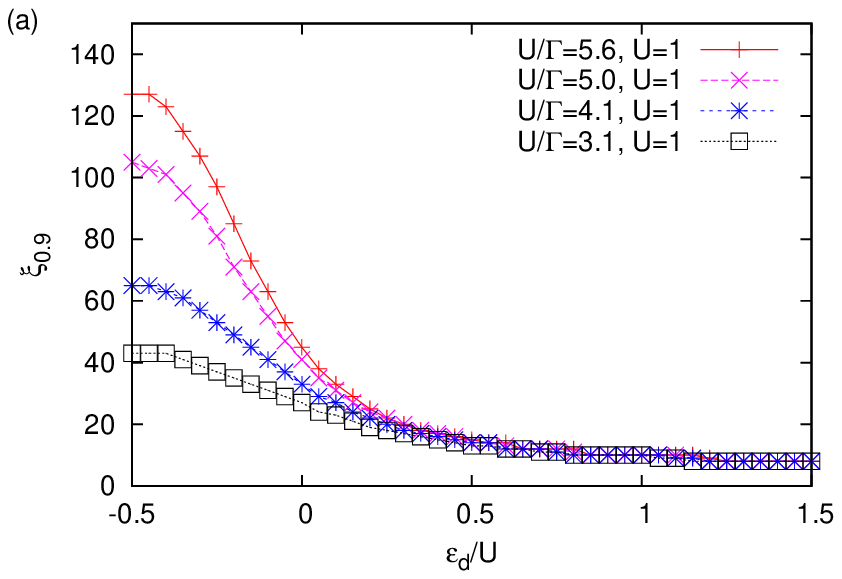}\\
  \includegraphics[width=\linewidth]{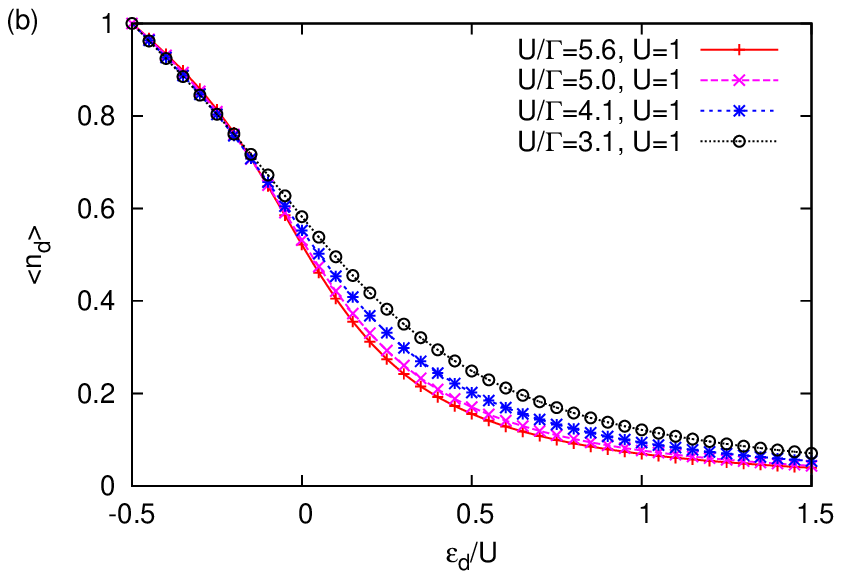}%
  \caption{(Color online) (a) Kondo screening length $\xi_{0.9}$ vs gate potential $\epsilon_d/U$ for several
    $U/\Gamma$ and $L=500$. (b) Dot occupation $\expect{n_{d}}$ vs gate potential.}
  \label{fig:gate-potential}
\end{figure}
We next investigate the behavior of the Kondo screening length while sweeping the gate potential applied to the dot.
Qualitatively, one expects the Kondo temperature to increase upon gating the dot away from particle-hole symmetry
and eventually, as the dot's charge starts to deviate substantially from one, the Kondo effect will be fully 
suppressed.\cite{Hewson1997} Consequently, we expect the Kondo cloud to shrink upon varying $\epsilon_d$. To elucidate 
this behavior, we focus on values of $U/\Gamma\lesssim 5.6$ for which $\xi_{0.9}(L=500)$ yields a good estimate of the 
true $\xi_K$, as  demonstrated in Sec.~\ref{sec:spin-correlations}.  

Our results for $\xi_{0.9}$ are presented in \figref{fig:gate-potential}(a). In addition, and as an illustration, we
plot the dot level occupation $\expect{n_{d}}=\bra{0}n_{d\shortuparrow} + n_{d\shortdownarrow}\ket{0}$ in
\figref{fig:gate-potential}(b), where $\ket{0}$ is the ground state of the system, obtained via DMRG. As we shift the
dot level away from the particle-hole symmetric point at $\epsilon_{d} = -{U}/{2}$ and thus leave the Kondo regime,
$\xi_{0.9}$ falls off rapidly. This is symmetric in the direction of the deviation from the Kondo point. In the regime
$\epsilon_{d} \lesssim -\Gamma$ one would expect Eq.~\eqref{eq:kondo-scale} to hold roughly. Indeed, for $\epsilon_{d} =
-U/4$, Eq.~\eqref{eq:kondo-scale} still applies,\footnote{Note that the prefactor $p$ depends on the gate potential,
  {\it i.e.}, $p=p(\epsilon_{d})$.} while for, {\it e.g.}, $\epsilon_{d}=0$ this is not the case anymore. The reason is
that Eq.~\eqref{eq:kondo-scale} is only valid in the Kondo regime with $\expect{n_{d}}\approx 1$. From
\figref{fig:gate-potential}(b) we see that the dot occupation starts to decrease quickly as we increase $\epsilon_{d}$
from $-U/2$, implying that the magnetic moment decreases as well. In the mixed-valence regime, $\epsilon_{d} \gtrsim
-\Gamma$, $\xi_{0.9}$ measures the strength of the spin-spin correlations not originating from Kondo physics.

\begin{figure}[ht!]
  \includegraphics[width=\linewidth]{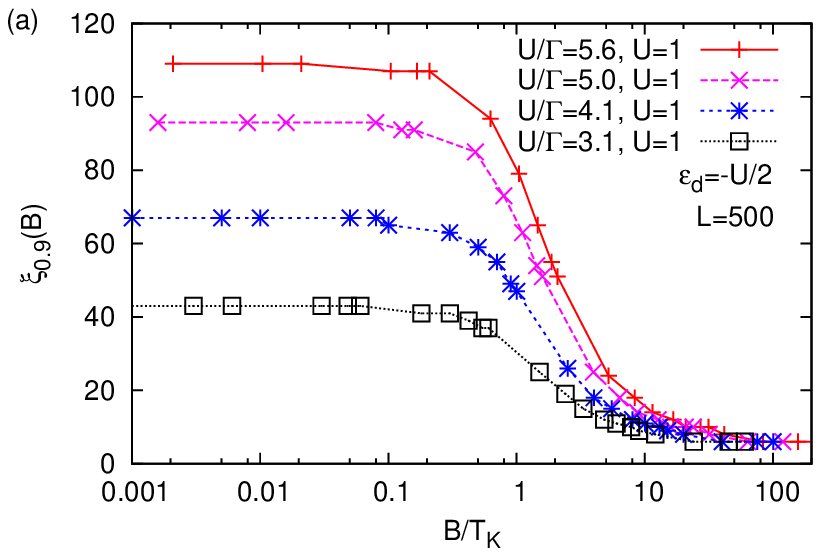}\\
  \includegraphics[width=\linewidth]{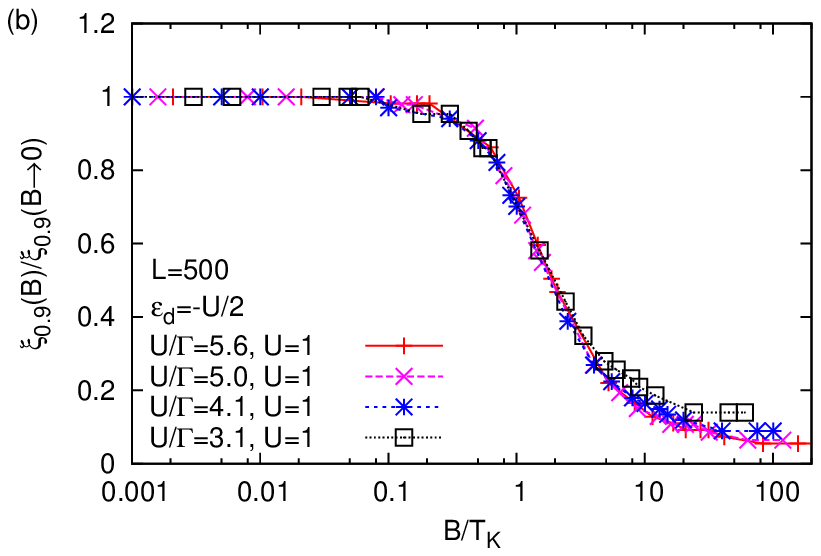}\\
  \includegraphics[width=\linewidth]{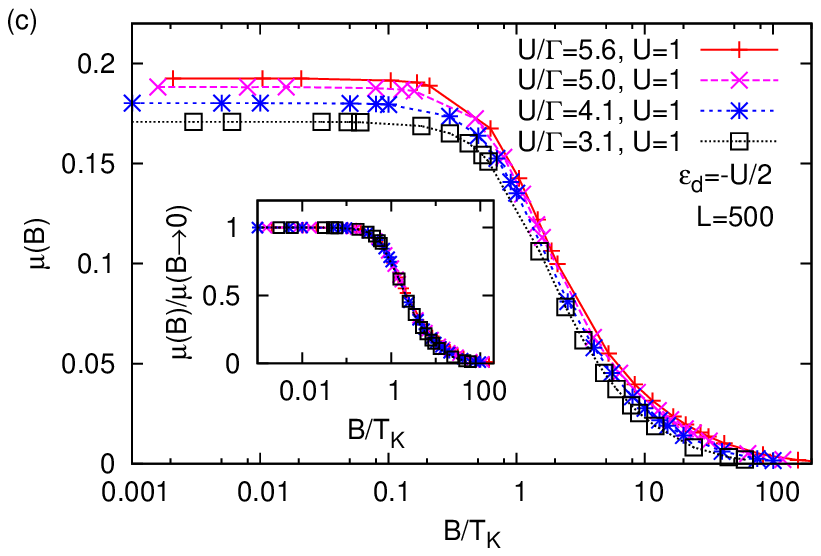}%
  \caption{(Color online) (a) Kondo screening length $\xi_{0.9}$ as a function of the magnetic field applied to the dot 
    for $L=500$. In all panels, $T_{K}$ is given by $T_{K} = \hbar v_{F}/\xi_{K}^{0}$ with $\xi_{K}^{0}$ from
      Eq.~\eqref{eq:kondo-scale}. (b) Scaling collapse of $\xi_{0.9}(B)/\xi_{0.9}(B\rightarrow 0)$ vs $B/T_{K}$ (c)
    Magnetic moment $\mu = \expect{(S_{d}^{z})^{2}} - \expect{S_{d}^{z}}^{2}$ vs. $B/T_{K}$. The inset shows the
    rescaled data $\mu(B)/\mu(B\rightarrow 0)$.}
  \label{fig:magnetic-field}
\end{figure}

\section{Magnetic field}
\label{sec:field}

The application of a magnetic field is known to destroy the Kondo effect and its influence on the density of states
(DOS) and the conductance has been widely studied.\cite{Costi2001,RoschCostiPaaskeWolfle2003} Here, we investigate how
the screening cloud collapses as the magnetic moment is squeezed by the magnetic field. In the presence of a finite
magnetic field the total spin $\vec{S}$ is no longer conserved but only $S^{z}$ is conserved. Thus we are left with  a
U(1) symmetry for $S^{z}$ instead of the SU(2) symmetry for $\vec{S}$. As a consequence, much more computational effort
is needed in order to achieve an accuracy similar to the zero-field case (see the Appendix for detail).

Our results for (i) the screening length $\xi_{0.9}(L=500)$ and (ii) the magnetic moment of the dot
$\mu=\expect{(S_{d}^{z})^{2}} - \expect{S_{d}^{z}}^{2}$ are displayed in Figs.~\ref{fig:magnetic-field}(a) and
\ref{fig:magnetic-field}(c), respectively. As the magnetic field is increased but still smaller than $T_{K}$, there are
almost no visible effects in $\xi_{0.9}$ (note the logarithmic scale in the figure). Once the magnetic field $B$ reaches
the order of the Kondo temperature $T_{K}$, the Kondo effect gets suppressed and the extent of the Kondo cloud shrinks
rapidly. More precisely, a pronounced decay of the screening length sets in at $B \simeq 0.5T_{K}$, in agreement with
findings for the field-induced splitting of the central peak in the impurity spectral function.\cite{Costi2000}
Qualitatively, both the screening length and the magnetic moment $\mu$ exhibit the same behavior. Note that for small
$U/\Gamma$, charge fluctuations reduce the magnetic moment to lie below the value $\mu = 1/4$ applicable for the Kondo
model, which presupposes $U/\Gamma \gg 1$.

To identify the point at which the Kondo effect breaks down, we again study the collapse of results from
\figref{fig:magnetic-field} onto a universal curve. This is shown in \figref{fig:magnetic-field}(b), and as a main
result we find:
\begin{equation}
  \label{eq:B-universal}
  \frac{\xi_{0.9}(B)}{\xi_{0.9}(B\rightarrow 0)} \propto f(B/T_K)\,,
\end{equation}
where $f(x)$ describes the universal dependence on $B/T_K$. We note that due to higher numerical effort for calculations
with a finite magnetic field (as further discussed in the Appendix) our numerical results slightly underestimate
$\xi_{0.9}(B)$ at $U/\Gamma\gtrsim 5$, in particular, at small $B$. This, however, has no qualitative influence on the
scaling collapse described by Eq.~\eqref{eq:B-universal}. We suggest that an analysis analogous to the one presented in
\figref{fig:magnetic-field} could be used to extract $T_{K}$ for models in which the dependence of $T_{K}$ on model
parameters is not known. In such an analysis, $T_K$ would be the only fitting parameter, since $\xi_{0.9}(B,L\to\infty)$
can be determined along the lines of Sec.~\ref{sec:spin-correlations} and one would obtain $T_{K}$ up to an unknown
prefactor, which is independent of $U/\Gamma$.

By rescaling the magnetic moment data to $\mu(B)/\mu(B\rightarrow 0)$ as shown in the inset of
\figref{fig:magnetic-field}(c) we again find a universal curve very similar to the  collapse of
$\xi_{0.9}(B)/\xi_{0.9}(B\rightarrow 0)$ in \figref{fig:magnetic-field}(b). We thus confirm that a collapse of local
quantities can be used to extract $T_{K}$, as previously shown using DMRG.\cite{SorensenAffleck1996} In principle, both
a scaling analysis of $\xi_{0.9}(B)$ and $\mu(B)$ can be used to extract $T_{K}$. Using the analysis of the screening
length data ($\xi_{K}$) offers the possibility of a scaling analysis as outlined in Sec.~\ref{sec:spin-correlations} to
reach parameter regimes, where a convergence of the data in $L$ has not yet been reached. Moreover, the analysis of
$\xi_K$ directly unveils the relevant length scales.

\section{Summary}
\label{sec:summary}

In this work, we studied the spin-spin correlations in the single-impurity Anderson impurity model using a
state-of-the-art implementation of the density matrix renormalization group method. We first considered the
particle-hole symmetric point and discussed two ways of collapsing the system-size-dependent data onto universal scaling
curves to extract a measure of the Kondo cloud's extension, the screening length $\xi_K$, as a function of
${U}/{\Gamma}$, or $T_K$, respectively. The first analysis is based on a scaling collapse of the integrated
correlations, while the second one employs a finite-size scaling analysis of the distance $\xi_a(L)$ from the impurity
at which a certain fraction $a$ of the impurity's magnetic moment is screened. $\xi_a(L)/\xi_a(\infty)$ exhibits a
universal dependence on $L/\xi_a(\infty)$, independently of the parameter $U/\Gamma$. We further showed that for an
appropriately chosen value of the parameter $a$, both approaches yield quantitatively similar estimates of the screening
length. Our results  for $\xi_K$, obtained from either of the scaling analyses, nicely follow the expected dependence on
${U}/{\Gamma}$.

As DMRG works in real space, the scaling regime could only be reached for ${U}/{\Gamma}=4$ and system sizes of
$L\lesssim 500$, but even for larger ${U}/{\Gamma}\lesssim 6$, a collapse onto the universal behavior could be achieved. Note
that ${U}/{\Gamma} \sim 4$ is the regime, in which time-dependent DMRG is able to capture Kondo correlations in real-time 
simulations of transport\cite{Al-HassaniehFeiguinRieraBuesserDagotto2006} on comparable system sizes, consistent
with our observations.

While NRG is better suited to access the regime of very small Kondo temperatures $T_K$, DMRG efficiently gives access to
the full correlation function $\expect{\vec{S}_{d}\cdot \vec{S}_{i}}$ in a single run. As an outlook onto future
applications, we emphasize that  DMRG allows for the calculation of the spin-spin correlations in the case of
interacting leads\cite{CostamagnaGazzaTorioRiera2006} or out-of-equilibrium, which is challenging if not impossible for
other numerical approaches with current numerical resources.

While the first part of our study focused on the particle-hole symmetric point where Kondo physics is dominant, we have further
analyzed how the screening cloud is affected (i) by varying the gate voltage and tuning the system into the mixed-valence regime,
and (ii) by applying a magnetic field at particle-hole symmetry. The latter provides an independent measure of the
Kondo temperature, through the universal dependence of the screening length on $T_K/B$.

{\it Note added:} while finalizing this work, we became aware of a related effort on the Kondo cloud,
Ref.~\onlinecite{BuesserMartinsRibeiroVernekAndaDagotto2009}, using the so-called embedded-cluster approximation, slave
bosons, and NRG. Their analysis is based on calculating the local density of states in the leads, as a function of the
distance from the impurity.

\begin{acknowledgments}
  We gratefully acknowledge fruitful discussions with  E.~Anda, L.~Borda, C.~B\"usser, E.~Dagotto, G.~B.~Martins,
  J.~Riera, and E.~Vernek. This work was supported by DFG (SFB 631, De-730/3-2, SFB-TR12, SPP 1285,
  De-730/4-1). Financial support by the Excellence Cluster ``Nanosystems Initiative Munich (NIM)'' is gratefully
  acknowledged. J.v.D and F.H.M. thank the KITP at UCSB, where this work was completed, for its hospitality. This
  research was supported in part by the National Science Foundation under Grant No. NSF PHY05-51164.
\end{acknowledgments}

\appendix

\section{Numerical Detail}
\label{sec:app}

\begin{figure}[t]
  \centerline{\includegraphics[width=0.9\linewidth]{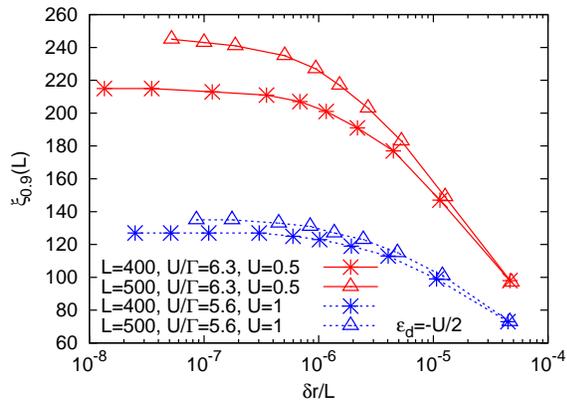}}
  \caption{(Color online) Convergence of $\xi_{0.9}$ vs. the residual norm per site $\delta r/L$ for
    $\epsilon_{d}=-U/2$, extracted from ground-state calculations using the SU(2) symmetry. For each combination of
    $U/\Gamma$ and $L$, the number of states kept increases for data points from right to left as $m = 200, 400, 600,
    800, 1000, 1200, 1500, 2000, 2500, 3000$, except for the case $U/\Gamma = 5.6$, $L = 500$, where no point with $m =3000$ is shown.}
  \label{fig:m-convergence}
\end{figure}

In this appendix we provide details on our numerical method. The DMRG calculations presented in this work are
challenging for two reasons. First, we model the conduction band with a chain of length $L$ that provides an energy
resolution of $1/L$, whereas the Kondo temperature becomes exponentially small with increasing $U/\Gamma$ [c.f.\
\Eq{eq:kondo-scale}]. Second, the spin-spin correlators are long-ranged quantities making very accurate calculations of
quantities necessary that are small compared to the unit of energy, $t$. The parameter controlling the accuracy of our
calculations is the number of states $m$ used to approximate the ground state during the DMRG sweeps. Typically, we
choose $m = 1500$ (3000 at most) for the calculation of the ground state. This results in  a residual norm per
site,\cite{verstraete:vmps} a measure for the quality of the convergence of the calculated ground state towards an
eigenstate of the Hamiltonian, $\delta r = \bra{\psi_{0}}(\hat{H}-E)^{2}\ket{\psi_{0}}$, on the order $\frac{\delta
  r}{L} = \mathcal{O}(10^{-7})$.

\begin{figure}[t]
  \centerline{\includegraphics[width=\linewidth]{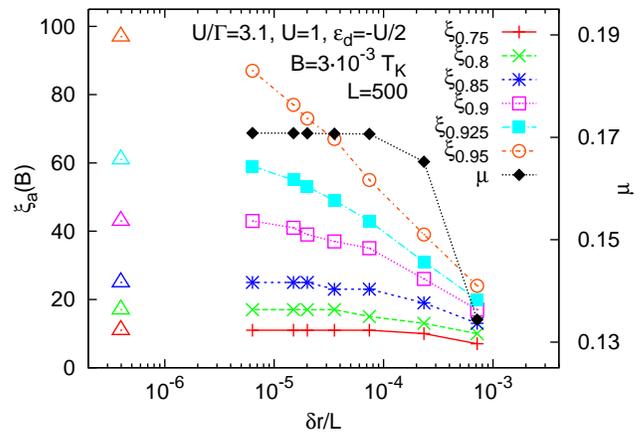}}
  \caption{(Color online) Convergence of $\xi_{a}(B)$ and the magnetic moment $\mu$ vs the residual norm per site
    $\delta r/L$ for a finite magnetic field of $B/T_{K}=3\cdot 10^{-3}$ and $\epsilon_{d}=-U/2$. For comparison, we add the
     $B=0$ data, represented by the open triangles, from calculations exploiting the SU(2) symmetry ($m=1500$ states kept). 
    The calculation with a magnetic field (symbols connected with lines) uses the U(1) symmetry only
    ($m=100,200,400,600,800,1000,1500$ states kept from right to left). The results for the dot's magnetic moment $\mu$
    are also included for comparison (solid diamonds).}
  \label{fig:conv-magn}
\end{figure}
Figure~\ref{fig:m-convergence} illustrates the $m$ dependence of $\xi_{0.9}$ for two values of $U/\Gamma$ and two values
of $L$ at $\epsilon_d=-U/2$, obtained from simulations using the SU(2) symmetry. The larger the ratio $U/\Gamma$ and the
bigger the system size $L$, the higher the number of states $m$, needed to be kept to obtain a well-converged ground
state, see \figref{fig:m-convergence}. This can be understood as follows: higher $U/\Gamma$ implies a smaller Kondo
temperature, {\it i.e.}, a larger screening length $\xi_{0.9}$ and longer-ranged spin-spin correlators
$\expect{\vec{S}_{d}\cdot\vec{S}_{i}}$. A well-converged ground state requires these to be evaluated accurately over the
entire range $i \lesssim \xi_{0.9}$, and hence more states need to be kept during the DMRG sweeps. For the scaling
analysis presented in Sec.~\ref{sec:spin-correlations} (see Figs.~\ref{fig:sigma-fit} and \ref{fig:xi-fit}), we only
used data points that are converged with respect to the number of states kept.

In \figref{fig:conv-magn}, we illustrate that the convergence with the number of states is greatly accelerated whenever
the SU(2) symmetry can be exploited. We compare this preferable case to  the calculations with a magnetic field, where
the SU(2) symmetry is reduced to a U(1) symmetry. In the figure, we use a small magnetic field of $B/T_{K}= 3\cdot
10^{-3}$, such that the results for $\xi_{a}(B,L=500)$ coincide with the results for $B=0$, previously obtained
from the SU(2) calculation. For instance, at $L=500$ by keeping $m=1500$ states, $\delta r \simeq 3\cdot 10^{-3}$ is
reached in the U(1) case as compared to $\delta r \simeq 2\cdot 10^{-4}$ for the SU(2) case. For $U=1$, $\Gamma=0.32$,
we show that this residual norm ensures accurate data for $\xi_{a}$ up to $a=0.9$, while for larger $a$, our U(1)
results are well below the corresponding SU(2) ones computed with the same $m$.

Pragmatically, in the case of broken SU(2) symmetry, one may resort to using a smaller threshold
$a$ (instead of $a=0.9$), for which the convergence with $m$ is faster. 
As we have shown in \figref{fig:xi-fit}, $\xi_{K}$ can be extracted from $\xi_{a}$ with $0.6 \leq a\leq 0.95$ 
up to a nonuniversal prefactor using the schemes discussed in Sec.~\ref{sec:spin-correlations}.

In contrast to the screening length, the calculation of the magnetic moment $\mu$, a local quantity, is much better behaved. 
Thus $\mu$ does not suffer much from the slower convergence of the U(1) calculation
and converges quickly to a high precision (displayed as  diamonds in \figref{fig:conv-magn}).

%%\input{paper.bbl}
%%\bibliography{../../paperdb/physik}

\end{document}